\documentclass[aps,prc,twocolumn,showpacs,superscriptaddress]{revtex4-1}
\usepackage{natbib}
\usepackage{graphicx}
\usepackage{color}
\usepackage{physics}
\usepackage{amsfonts}
\begin{document}
\bibliographystyle{revtex}
\title{Spectroscopic properties of  $^{4}$He within a multiphonon approach}
\vspace{0.5cm}
\author{G. De Gregorio}
\affiliation{Dipartimento di Matematica e Fisica, Universit$\grave{a}$ degli Studi della Campania "Luigi Vanvitelli",
viale Abramo Lincoln 5, I-81100 Caserta, Italy}
\affiliation{Istituto Nazionale di Fisica Nucleare, Complesso Universitario di Monte S. Angelo, Via Cintia, I-80126 Napoli, Italy}
\author{F. Knapp}   
\affiliation{Institute of Particle and Nuclear Physics, Faculty of Mathematics and Physics,  Charles University, V Hole\v sovi\v ck\'ach 2, 180 00 Prague, Czech Republic } 
\author{N. Lo Iudice}
\affiliation{Dipartimento di Fisica, 
Universit$\grave{a}$ di Napoli Federico II, 80126 Napoli, Italy} 
\author{P. Vesel\'y} 
\affiliation{Nuclear Physics Institute,
Czech Academy of Sciences, 250 68 \v Re\v z, Czech Republic} 
\date{\today}

\begin{abstract}
Bulk and spectroscopic properties of $^4$He are studied within  an equation of motion phonon method.  Such a method generates a basis
of $n$-phonon  ($n=0,1,2,3...$) states composed of tensor products of particle-hole Tamm-Dancoff  phonons and then
solves the full eigenvalue problem in such a basis. The method does not rely on any approximation and is free of any contamination induced by the center of mass, in virtue of a procedure exploiting
the singular value decomposition of rectangular matrices.
Two potentials,  both derived from the chiral effective field theory, are adopted in a self-consistent calculation performed within a space including up to three phonons. The latter  basis states are treated under a simplifying assumption.
A comparative analysis with the experimental data points out the different performances of the two potentials. It shows also that the calculation succeeds only partially in the description of the spectroscopic properties
and suggests a recipe for further improvements.
\end{abstract}
\maketitle

\section{Introduction} 
The studies of few-body nuclear systems  have advanced rapidly in the past two decades. 
The   growing computational resources, combined with  highly efficient numerical algorithms, have enhanced greatly the performance  of traditional methods and stimulated the development of new techniques  (see Refs. \cite{LeidOrl13,Carlson2015}  for review and references). 

Most approaches adopt intrinsic coordinates  and therefore avoid any interference with the center of mass (c.m.) motion which  cannot be averted within standard  shell model (SM).
The  decoupling of the intrinsic from the c.m. motion  has been achieved within the no-core SM (NCSM)  (see Refs. \cite{Navra09,Barrett13} for a review and references) 
 under the following stringent conditions:
(i) Add a c.m. Hamiltonian of frequency $\omega$ to the intrinsic one  according to the Lawson prescription \cite{Gloekner74}, (ii) use a SM basis built of harmonic oscillator (HO) single particle (s.p.) states of the same frequency, and (iii) include all and only the configurations up to N$_{\rm max} \hbar \omega$.
Being bound to the HO SM basis, the method  can be formulated also in terms of Jacobi coordinates   \cite{Navratil1999}.  

The search for realistic nucleon-nucleon ($NN$) interactions has evolved in parallel with the search for reliable many body approaches. To this purpose methods for deriving from them effective interactions by softening their repulsive short range components have been developed.
Two notable effective potentials are the V$_{lowk}$ \cite{Nogga2004} obtained by integrating out the high-momentum components of the $NN$ interaction in the free space and the  correlated V$_{UCOM}$ obtained through the unitary correlation operator method (UCOM) \cite{Feldmeier1998}.  

It came out that the NN  interaction, if used alone, cannot describe the physics of the three-nucleon systems \cite{Kievsky08}. One needs to introduce the $NNN$ forces. Several semi-phenomenological $NNN$ interactions have been proposed  (see Ref. \cite{Carlson2015} for review and references). 
A more consistent scheme for their derivation  is provided by the chiral effective field theory ($\chi$EFT)~ \cite{Weinberg1990,Weinberg1991}, where the Hamiltonian is generated  as a series expansion in terms of the momentum or pion mass. The power counting introduces naturally $NN$, $NNN$, and higher order  interactions according to a specific hierarchy \cite{Epelbaum2009,Machl11}.  The $NNN$ forces appear already at third order (N2LO).  Recently, chiral $NN$ potentials incorporating all contributions from leading order (LO) up to fifth order (N4LO) have been determined with high accuracy \cite{Entem2017,Reinert2018}.
The $\chi$EFT interaction is often smoothed  through a similarity renormalization group
(SRG) transformation \cite{Bogner2007}  and, so renormalized, is suitable for calculations in truncated shell model spaces, thereby enlarging considerably the domain of applicability of  {\it ab initio} investigations (see Ref. \cite{Stroberg2019} for review and references). 

Several {\sl ab initio} approaches, mostly built on NCSM, have adopted the  N2LO or the N3LO potentials (see Ref. \cite{Navratil2016}  for review and references). Some of them have included the ground-state properties of $^4$He in the protocol adopted to determine the low-energy constants (LEC) of the $\chi$EFT potential  \cite{Ekstr15}. Other realistic calculations have been performed by resorting to the equations of motion method which is known to be an efficient tool for solving the nuclear eigenvalue problem.  We mention  the coupled cluster (CC) \cite{Hagen2014} and the in medium SRG (IMSRG) \cite{Hergert16}, and the random phase approximation (RPA) calculations using UCOM \cite{Paar2006} and NNLO$_{\rm{sat}}$ \cite{Parz2017,Wu2018}.  

A few years ago, we  developed for closed shell nuclei an equation of motion phonon method (EMPM)  \cite{AndLo,AndLo1,Bianco},
which yields an orthonormal multiphonon basis built out of phonons generated in the particle-hole (p-h) Tamm-Dancoff approximation (TDA) and adopts such a basis to solve the full eigenvalue problem under no approximation apart from the truncation of the multiphonon space.  The method was also formulated in the quasiparticle language suitable for open shell nuclei \cite{DeGreg16}  and  in the p(h)-phonon scheme for the study of odd-nuclei \cite{DeGreg16a,DeGreg17a,DeGreg17b,DeGreg18,DeGreg19,DeGregorio2020}.

Very recently, we have endowed the EMPM with a procedure  which removes the c.m. spurious admixtures under no constraint and for any s.p. basis \cite{DEGREGORIO2021}. It proceeds in two-steps. We first decouple the c.m.   from the  TDA states \cite{Bianco14} by exploiting the Gram-Schmidt orthogonalization method.  We then  remove the residual contaminations from  the multiphonon basis states by resorting to the singular value decomposition (SVD).This procedure can be easily extended to  odd  systems and can be adopted within  the quasiparticle EMPM to remove at once the contamination induced by the c.m. and    the violation of the particle number in open-shell nuclei. In order to show that the method so reformulated can be applied to very light nuclei  we have performed an exact calculation  using a Hartree-Fock (HF) basis derived from a restricted HO space (up to N$_{max}$=5) for $^4$He.  

Here we consider  a larger HO space in order to offer a more extensive and exhaustive investigation. $^4$He  was already  studied in approaches using Jacobi and hyperspherical coordinates (see for example Refs. \cite{Kievsky08,LeidOrl13}) as well as in NCSM \cite{Navratil07,Gazit09,Navra09,Jurgen2009,Jurgen2011},  CC \cite{Ekstr15,Bacca13}, IMSRG \cite{Hergert13a}, and RPA \cite{Wu2018}.  However, all mentioned studies were focused on its bulk properties or the giant resonance apart from an early NCSM  evaluation of the spectrum \cite{Zheng94} and a variational approach based on correlated Gaussians which uses the  Argonne v8' potential plus a phenomenological three-body force \cite{Horiuchi13}. 
 
Ours is a self-consistent approach which covers ground as well as excited states. We adopt two potentials,  NNLO$_{\rm{sat}}$ \cite{Ekstr15} and Daejeon16  \cite{Shirok2016}, both rooted in the effective field theory.  We will refer to them as V$_S$ and V$_D$, respectively. We will analyze  for the two potentials the convergence properties of the ground state (g.s.) observables with respect to the HO frequencies and space dimensions   as well as the convergence of two spectra  versus the HO frequencies. We also  establish an appropriate correspondence between computed and experimental levels by relating their decay mode to the phonon and p-h content of the computed states.  Finally, we will investigate the evolution of the electric dipole $E1$ strength distribution as the HO space dimensions increase and show how the redistribution of the $E1$ peaks affects the giant resonance (GR) cross section. We  hope that the present investigation may provide some useful insights on the structure of  $^4$He and  some probative indications for more refined tests of the available potentials. 
 
\section{ Brief outline of the method}
The basic ingredients are the HF p-h vacuum $\ket{0}$ and the TDA states
$\ket{\lambda} = O^\dagger_\lambda \ket{0}$ of energy $E_\lambda$,
where
\begin{equation}
\label{lam}
 O^\dagger_\lambda  = \sum_{ph} c^\lambda_{ph} (a^\dagger_p  \times b_h)^\lambda
\end{equation}
is the phonon operator built out of the creation and annihilation operators
 $a^\dagger_p=  a^\dagger_{x_p j_p m_p} $  and $b_h = (-)^{j_h + m_h} a_{x_h j_h - m_h}$, which are coupled to spin $J_\lambda$ and create p-h configurations of energy $\epsilon_p - \epsilon_h$. The ? denotes angular-momentum coupling.
 
Starting  from $\ket{0}$ and the TDA one-phonon states $\ket{\lambda}$, we intend to generate iteratively an  orthonormal  basis of $n$-phonon ($n=2,3..$) correlated states $\ket{\beta}  = \ket{\alpha_{n}}$ assuming known the  $(n-1)$-phonon basis states $\ket{\alpha} = \ket{\alpha_{n-1}}$ of energy $E_\alpha$. 
To this purpose we construct the set of redundant states 
 \begin{equation}
 \label{Olam}
 \ket{(\lambda \times \alpha)^{\beta}}=  \Bigl\{O^\dagger_\lambda \times \ket{\alpha} \Bigr\}^{\beta} 
 \end{equation}
and extract from them a basis of linearly independent states by resorting to the Cholesky decomposition method. We are then allowed to write the $n$-phonon states we search for in the expanded form  
\begin{equation}
  \ket{\beta} = \sum_{\lambda \alpha}  C_{\lambda \alpha }^{\beta} \ket{(\lambda \times \alpha)^{\beta}}.
\label{nstatec}
\end{equation}
They can be determined by solving the generalized eigenvalue equation within the $n$-phonon subspace 
\begin{equation}
\bra{(\lambda \times \alpha)^{\beta}} H \ket{\beta} =  E_\beta
\bra{(\lambda \times \alpha)^{\beta}}\ket{\beta}   
\end{equation}
For our purposes, however, it is more useful to exploit the structure (\ref{Olam}) of the states $\ket{(\lambda \times \alpha)^{\beta}}$ and use the equivalent equation of motion in the reduced form  
\begin{equation}
\langle \beta \parallel [H,O^\dagger_\lambda] \parallel \alpha \rangle =  
(E_{\beta} - E_\alpha) \langle \beta \parallel O^\dagger_\lambda \parallel{\alpha} \rangle.  
\label{EoM}
\end{equation}
Once expanded, the commutator contributes through terms like  
$\langle \beta \parallel [(a^\dagger_p \times b_h)^{\lambda'} \times (a^\dagger_r \times b_s)^\sigma]^{\lambda} \parallel \alpha \rangle $,  where $(rs)= (pp')$ and $(rs)=(hh')$. 
We then need just to act on these matrix elements by using the closure 
\begin{equation}
I_{n-1}= \sum_\alpha \dyad{\alpha} 
\end{equation}
 and  expressing the p-h operators $(a^\dagger_p \times b_h)^{\lambda'}$ in terms of $O^\dagger_ {\lambda'}$ upon  inversion of Eq. (\ref{lam}).

These operations lead to the generalized eigenvalue equation within the $n$-phonon subspace
\begin{equation}
\label{eig}
 {\cal H} C = ({\cal A} {\cal D}) C= E {\cal D} C,
\end{equation}
or, more explicitly,
\begin{equation}
\sum_{\lambda' \alpha'} {\cal H}^\beta_{\lambda \alpha \lambda' \alpha'} 
C^\beta_{\lambda' \alpha'} =  E_\beta \sum_{\lambda' \alpha'} {\cal D}^\beta_{\lambda \alpha \lambda' \alpha'}  C^\beta_{\lambda' \alpha'}. 
\label{eig1}
\end{equation}
Here
\begin{equation}
\label{D}
 {\cal D}^{\beta}_{\lambda \alpha \lambda' \alpha'}= \bra{(\lambda  \times \alpha )^{\beta}}\ket{(\lambda' \times \alpha')^{\beta}}
\end{equation}
is the overlap or metric matrix which preserves the Pauli principle and
\begin{equation}
{\cal H}^\beta_{\lambda \alpha \lambda' \alpha'} = 
\sum_{\lambda'' \alpha'' } {\cal A}^\beta_{\lambda \alpha \lambda'' \alpha''} 
{\cal D}^\beta_{\lambda''  \alpha''  \lambda' \alpha'},
\label{H1}
\end{equation}
where
\begin{eqnarray}
 {\cal A}^\beta_{ \lambda \alpha   \lambda'' \alpha''} =  (E_\lambda + E_{\alpha} ) \delta_{\lambda \lambda''} \delta_{\alpha \alpha''} 
+ {\cal V}^\beta_{\lambda \alpha \lambda'' \alpha''}.
\label{A}
\end{eqnarray}
The expressions of the overlap matrix ${\cal D}$ and of the phonon-phonon interaction  ${\cal V}$  can be found, for instance, in Ref. \cite{Bianco12}.
The solution of Eq. (\ref{eig1}) yields the $n$-phonon basis states (\ref{nstatec}).
    
 The iteration of the procedure up to an arbitrary $n$ produces a set of states which,  added to   HF ($\ket{0}$) and TDA   ($\{\ket{\alpha_1}\}= \{\ket{\lambda}\}$), form an orthonormal basis $\{ \ket{\alpha_n}\}$ ($n=0,1,2,3,...$). 
 Such a basis is then adopted to solve the eigenvalue problem  in the full space
\begin{equation}
\label{eigfull}
\sum_{ \alpha_n \beta_{n'}} \Bigl((E_{\alpha_n} - {\cal E}_\nu) \delta_{\alpha_n \beta_{n'}}  +   {\cal V}_{\alpha_n \beta_{n'}} \Bigr){\cal C}^{\nu}_{\beta_{n'}} = 0,
\end{equation}
where ${\cal V}_{\alpha_n \beta_{n'}}  = 0$ for  $n' = n$.

For $n'=n+1$ ($n>0)$ we have  ($\alpha = \alpha_n, \beta = \beta_{n +1}$)
\begin{equation}
\label{couponen}
{\cal V}_{\alpha \beta} =     
   \sum_{ \sigma \alpha'}    {\cal V}^\sigma_{\alpha \alpha'}  \langle(\sigma \times \alpha')^\beta \mid \beta \rangle, 
\end{equation}
where
\begin{equation}
{\cal V}^\sigma_{\alpha \alpha'}   = 
\frac{1 }{[\alpha ]^{1/2}}  (-)^{\alpha + \alpha' + \sigma}    
   \sum_{r \leq s}  {\cal V}^\sigma_{r s} 
	\langle \alpha' \parallel \bigl(a^\dagger_r \times b_s \bigr)^\sigma \parallel \alpha \rangle  
\end{equation} 
and
\begin{eqnarray}
{\cal V}^\sigma_{rs}= \sum_{p h }  c^\sigma_{ph}  F^\sigma_{r s p h}. 
\end{eqnarray}
Here 
\begin{equation}
F^\sigma_{r s p h} = \sum_\gamma (2 \gamma + 1) (-)^{r + h - \sigma - \gamma}
W(rpsh;\gamma \sigma) V^\gamma_{rpsh},
\end{equation}
where $V$ is  the two-body potential and $W$ are Racah coefficients.

The coupling of the  vacuum  to the  two-phonon states   is given by
\begin{equation}
\bra{0} H \ket{\alpha_{2}} = \sum_{\lambda \lambda'} \bra{(\lambda \times  \lambda')^0}\ket{\alpha_2} \bra{\lambda} V \ket{\lambda'}.
\label{couptwo}
\end{equation}
For $n' = n + 2$ ($n > 0$) the matrix elements can be written in the simple form  
\begin{equation}
{\cal V}_{\alpha_n \beta_{n'}} = \sum_{\alpha_2} \bra{0} H \ket{\alpha_2} 
\bra{(\alpha_2 \times \alpha_n)^\beta}\ket{\beta_{n'}} 
\label{couptwon}
\end{equation}   
The solution of the final eigenvalue Eq. (\ref{eigfull}) yields the eigenvectors ($n=0,1,2,3....$)
\begin{equation}
\label{Psifull}
\ket{\Psi_\nu} = \sum_{n,\alpha_n}  {\cal C}_{\alpha_n}^\nu \ket{\alpha_n}.
\end{equation}

\section{Removal of the c.m. motion} 
The preliminary  step \cite{Bianco14} consists in adopting Gram-Schmidt to extract from  the $n_{ph}$ p-h configurations a set of $n_{ph}-1$ states orthogonal to
\begin{equation}
\ket{\lambda_{c.m.}, \mu} = \frac{1}{N} R_\mu \ket{0},
\label{TDAcm}
\end{equation}
where $R$ defines the c.m. coordinates and $N$  is a normalization constant.
The states so obtained yield $n_{ph}-1$ c.m. free TDA phonons.
 
Let us now consider the two-phonon subspace and separate the $n$ states 
\begin{equation}
\{\ket{i}\} = \{\ket{(\lambda \times \lambda')^\alpha}\},
\end{equation} 
composed of the c.m. free phonons $\ket{\lambda}$,
from the $m$ ones 
\begin{equation}
\{ \ket{s} \} = \{\ket{(\lambda \times \lambda_{c.m.})^\alpha} , \ket{(\lambda_{c.m.} \times \lambda_{c.m.})^\alpha} \} 
\end{equation} 
containing at least one c.m. phonon $\ket{ \lambda_{c.m.}} $.
The overlap between the two set of states  is non vanishing
 \begin{equation}
  {\cal D}^{(c.m.)}_{si} = \bra{s}\ket{i} \neq 0
 \end{equation}
and, therefore, reintroduces the c.m. contamination  in the two-phonon states $\ket{\alpha}$.
    
We need, therefore, to construct a new basis 
of states
\begin{equation}
\ket{\alpha} = \sum_i C^{\alpha}_i \ket{i}   
\end{equation} 
out of the set of $\ket{i}$, which fulfills the orthogonality condition   
\begin{equation}
\bra{\alpha}\ket{s} =0 
\end{equation}
for all $\ket{s}$ c.m. states and any $\ket{\alpha}$.
This amounts to determine the right null space of the rectangular matrix ${\cal D}^{(c.m.)}$
\begin{equation}
{\cal D}^{(c.m.)} C =0,
\label{Dcm0}
\end{equation}
a goal achieved by  a procedure exploiting the SVD.

According to the SVD, the $m \times n$ rectangular matrix  $ {\cal D}^{(c.m.)}$ undergoes the following decomposition  
\begin{equation}
{\cal D}^{(c.m.)} = U  \Sigma V^T = \sum_{i=1,m} {\mathbf u}_i \sigma_i  {\mathbf v}_i,
\label{Dcm}
\end{equation}
 where $U$ is a left-singular orthonormal $m \times m$ matrix composed of the row singular vectors ${\mathbf u}_i $ acting on the c.m. states,
 $V^T$ is the transpose of a right-singular orthonormal $n \times n$ matrix $V$ composed of the column singular vectors ${\mathbf v}_i$ acting on the states composed of c.m. free  phonons $\ket{\lambda}$, and $\Sigma$ is an $m \times n$ rectangular diagonal matrix with $m$ non vanishing singular values $\sigma_i \neq 0$.

It is to be noted that the other $n-m$ singular values vanish,  $\sigma_i =0$ for $i= m+1, n$. 
 Thus,  the  right-singular matrix $ V$  decomposes into two submatrices.  One is composed of the vectors ${\mathbf v}_s$  ($s=1,m$) and  yields the transformed   states
\begin{equation} 
\ket{\nu_s} = \sum_{i= 1,n} v_{si} \ket{i} \hspace{1cm} (s= 1 , m)
\end{equation}
spanning the c.m. spurious subspace.

The other submatrix, which we denote by  $\mathbb{V}$, is composed of the singular vectors ${\mathbf v}_r$  ($r=m+1,n$) and generates the $n-m$ transformed  states
\begin{equation} 
\ket {\nu_r}  =  \sum_{i= 1,n} v_{ri} \ket{i}  \hspace{1cm} (r= m+1 , n)
\end{equation}
orthogonal to the c.m. states $\ket{\nu_s}$
\begin{equation}
\bra{\nu_r}\ket{\nu_s} =0. 
\end{equation}
These states form the intrinsic subspace we searched for. 

We can then apply  the transformation $\mathbb{V}$ to the eigenvalue Eq. (\ref{eig}) obtaining
\begin{equation}
 {\cal H}' C' = E {\cal D}'  C', 
\label{eigfree}
\end{equation}
where $C' = \mathbb{V} C$, ${\cal D}'= \mathbb{V} {\cal D} \mathbb{V}^T$, and ${\cal H}'= \mathbb{V} {\cal H} \mathbb{V}^T$.

The eigenstates  
\begin{equation}
\ket{\alpha_2} = \sum_{r} (\mathbb{V} C)_{r} \ket{\nu_r} 
\end{equation}
can be recast in terms of the original basis states $\ket{i} = \ket{ (\lambda \times \lambda')^\alpha }$.

We adopt the same procedure for the three-phonon subspace once we   identify the c.m. spurious states.
These are  
\begin{equation}
\ket{s} = \{\ket{(\lambda_{cm} \times \alpha)^\beta} ,\ket{(\lambda \times \alpha_{cm})^\beta} , \ket{(\lambda_{cm} \times \alpha_{cm})^\beta} \},
\end{equation}
 where $\ket{\alpha_{cm}}$ are just the transformed two-phonon states $\ket{\nu_s}$  ($s=1,m$)  corresponding to the non vanishing singular values $\sigma_s \neq 0$.
  
\section{Numerical implementation and results}
We adopt a Hamiltonian of the form
\begin{equation}
H = T_{int} + V,
\end{equation}
where $T_{int}$ is the intrinsic kinetic energy and  V is either V$_S$ \cite{Ekstr15} or  V$_D$ \cite{Shirok2016}.   
V$_S$ is obtained by optimizing simultaneously the two-body and three-body components of the $\chi$EFT potential at  N2LO with a cutoff parameter $\Lambda=450$ MeV. In the present calculation the full three-body force is used to generate the HF basis and is truncated at the normal ordered two-body level in solving the multiphonon eigenvalue problem.
V$_D$ is derived from  the $NN$ component of the N3LO potential \cite{EndMach03} in two steps.  The $NN$ potential  is first softened by a SRG method \cite{Wegner94}  with flow parameter $\lambda=1.5$ fm$^{-1}$  and then  subjected to a phase equivalent transformation which determines an optimal set of parameters of the $NN$ force. The absence of  three-body forces reduces considerably the computational effort.

The numerical procedure goes through the following steps: (i) Derive a HF basis  from a HO space of dimensions  N$_{max}$  and frequency  $\omega$.; (ii) use the HF states to create the TDA phonon basis; (iii)generate  the $n$-phonon ($n=2,3..$) basis by deriving and solving iteratively   the EMPM Eq. (\ref{eig}); and (iv) the basis so constructed is adopted to solve the final eigenvalue problem in the multiphonon space [Eq. (\ref{eigfull})].

We have performed already an exact calculation  using the full $n$-phonon basis up to $n=3$ within a HO space including six major shells (N$_{\rm{max}}$=5). The results obtained using V$_S$ were presented elsewhere \cite{DEGREGORIO2021}.

For the larger spaces (N$_{max} > 5$) considered here,  we solve exactly the full eigenvalue problem  [Eqs. (\ref{eig}) and (\ref{eigfull})]  up to two phonons ($n=2$) .
An exact treatment for $n > 2$ would be too time-consuming. On the other hand,  the three-phonon states are far above the experimental region and affect
the low-lying one-phonon and two-phonon states only through their coupling.   Therefore, we treat them in the diagonal approximation.   
Namely, we neglect the phonon-phonon interaction ${\cal V}_{\lambda \alpha \lambda' \alpha'}$  in Eq. (\ref{A}) so that the three-phonon eigenvalues are simply
\begin{equation}
\label{E3diag}
E_{\beta_3} \simeq E_{\alpha_2} +  E_\lambda. 
\end{equation}
\begin{figure}[ht!]
\includegraphics[width=\columnwidth]{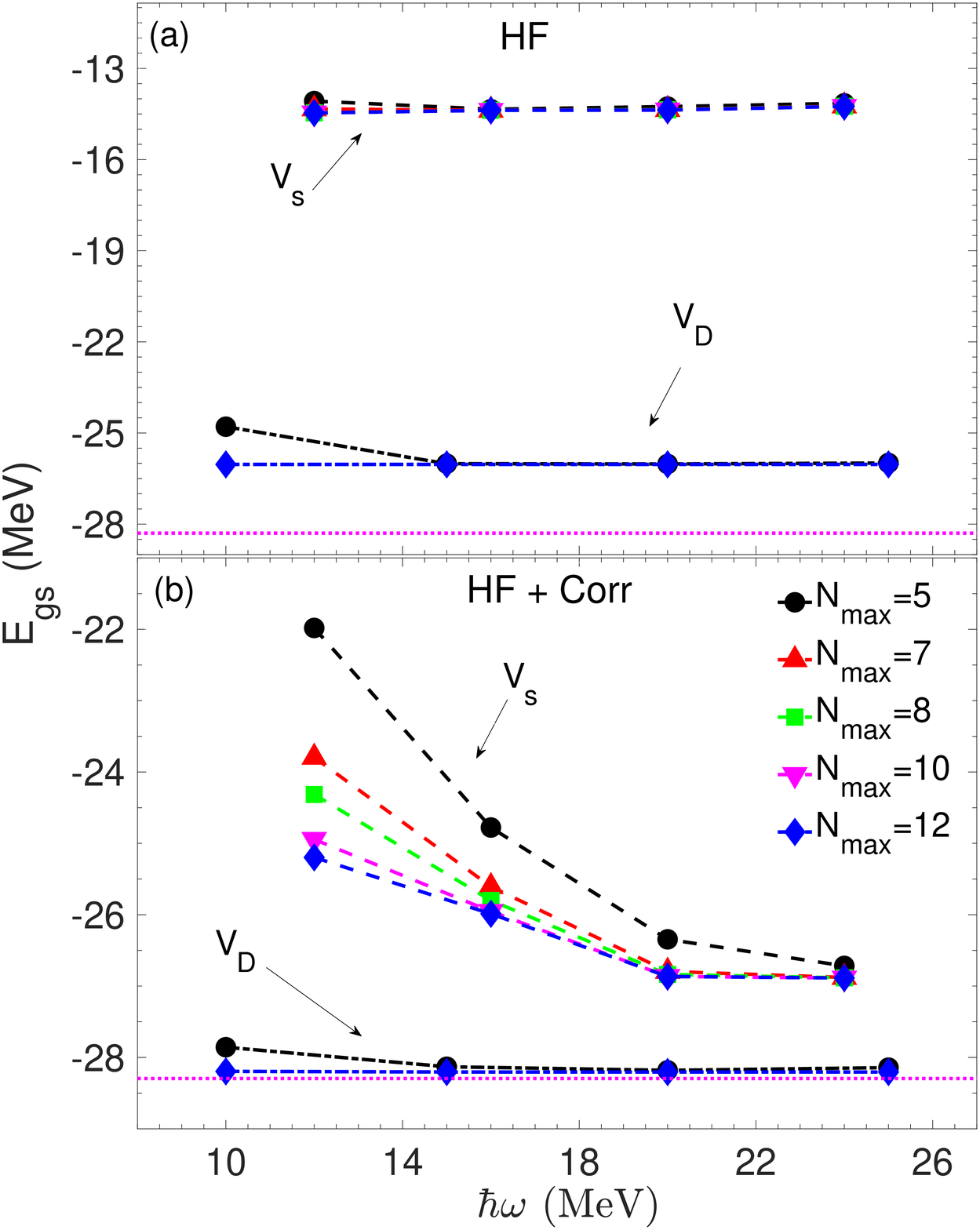}
\caption{(Color online) \label{fig1} HF and total g.s. energies produced by  V$_S$ (dashed lines) and V$_D$ (dot-dashed lines) versus  the HO frequency  for different dimensions N$_{\rm max}$. The dotted line indicates the experimental value \cite{Angeli13}.} 
\end{figure}

\begin{figure}[ht!]
\includegraphics[width=\columnwidth]{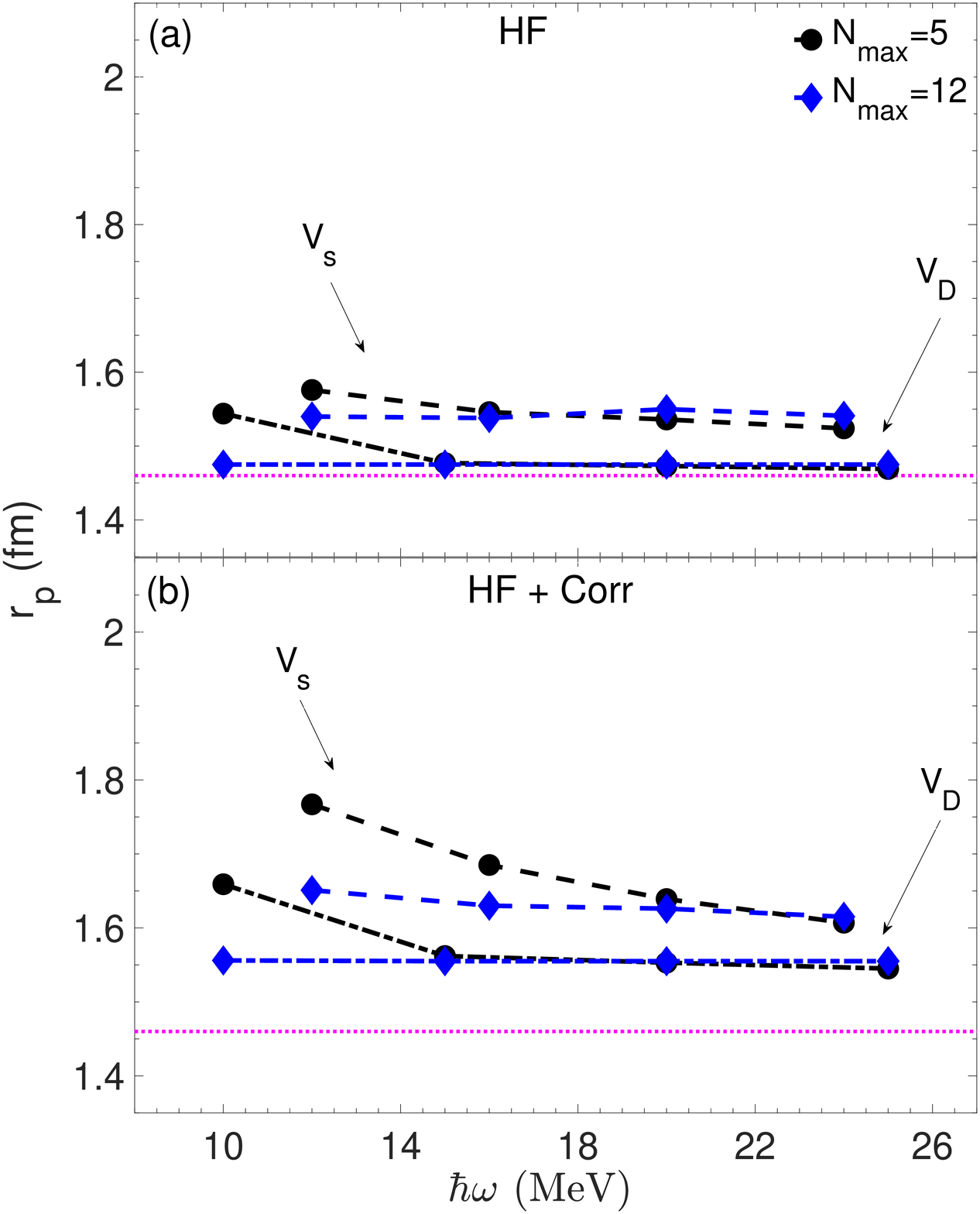}
\caption{(Color online) \label{fig2} V$_S$ (dashed lines) and V$_D$ (dot-dashed lines) HF and total proton radii $r_p$ versus the HO frequency    for different dimensions N$_{\rm max}$. The dotted line indicates the experimental value \cite{Angeli13}.}
\end{figure}

\begin{figure}[ht]
\includegraphics[width=\columnwidth]{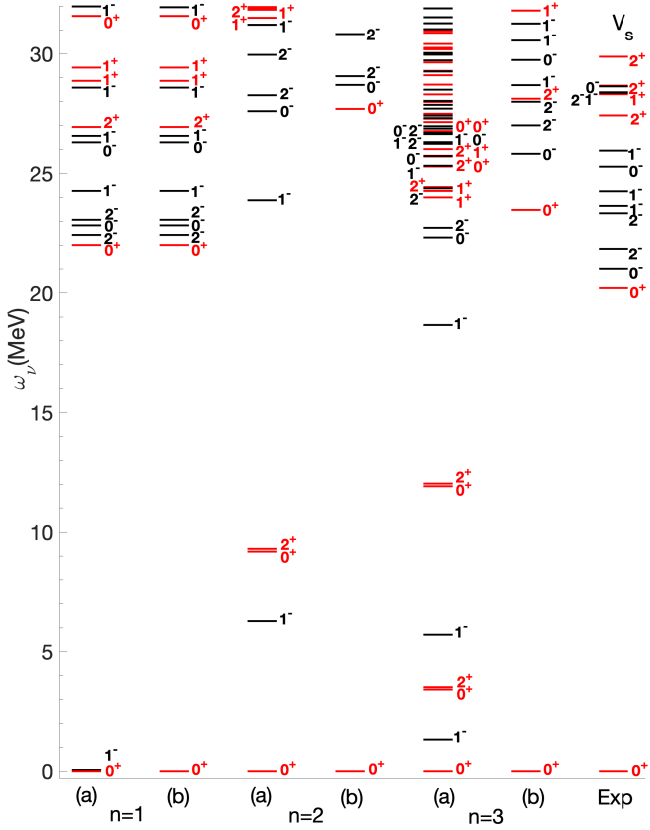}
\caption{(Color online) \label{figcm} Spectra obtained in different multiphonon spaces before (a) and after (b) having removed the c.m. motion.  The calculation was performed using V$_S$ within the restricted N$_{max}$ =5  HO space for $\hbar \omega =20$ MeV.  The levels are referred to the HF g.s. ($\omega_\nu = E_\nu - E_{HF}$)  in TDA (n=1) and to 
the corrrelated g.s. ($\omega_\nu = E_\nu - E_0$) for $n=2$ and $n=3$. The experimental data are from Ref. \cite{NNDC}. }
\end{figure}

\begin{figure}[ht]
\includegraphics[width=\columnwidth]{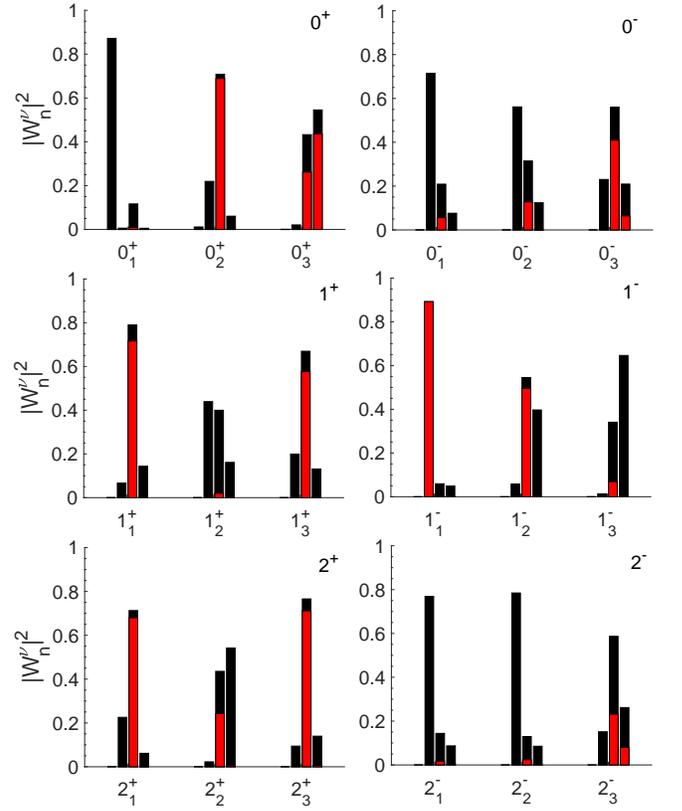}
\caption{(Color online) \label{figwf} Physical (black) and c.m. spurious (red) content of the different $n$-phonon components of some typical states before SVD. $W^{\nu}_n=\sum_{\alpha_n} \mid {\cal C}^{\nu}_{\alpha_n}\mid^2$ gives the total weight of the different $n$-phonon components for a given $n$ [see Eq. (\ref{Psifull})]. The bars are ordered from left to right following the sequences $n=0,1,2,3$ for the 0$^+$ states  and $n=1,2,3$ for the others.}
\end{figure}

\begin{figure}[ht]
\includegraphics[width=\columnwidth]{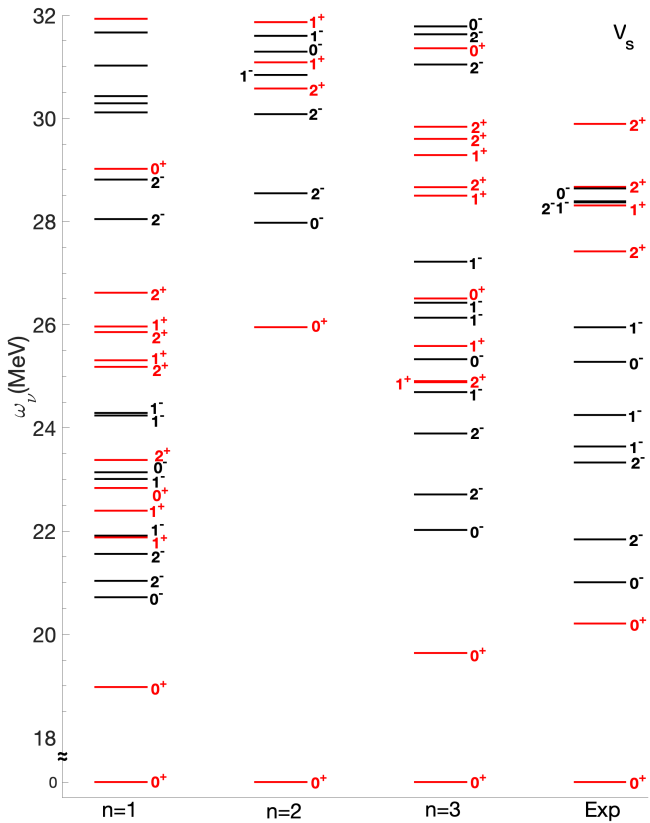}
\caption{(Color online) \label{fig3} Spectra computed using  V$_S$ for $\hbar \omega =20$ MeV and N$_{max}$ = 12 for different multiphonon spaces ($n=1,2,3$). The levels are referred to the HF g.s. 
($\omega_\nu = E_\nu - E_{HF}$)  in TDA ($n=1$) and to the correlated g.s. ($\omega_\nu = E_\nu - E_0$) for $n=2$ and $n=3$. The experimental data are from Ref. \cite{NNDC}. }
\end{figure}

\begin{figure}[ht]
\includegraphics[width=\columnwidth]{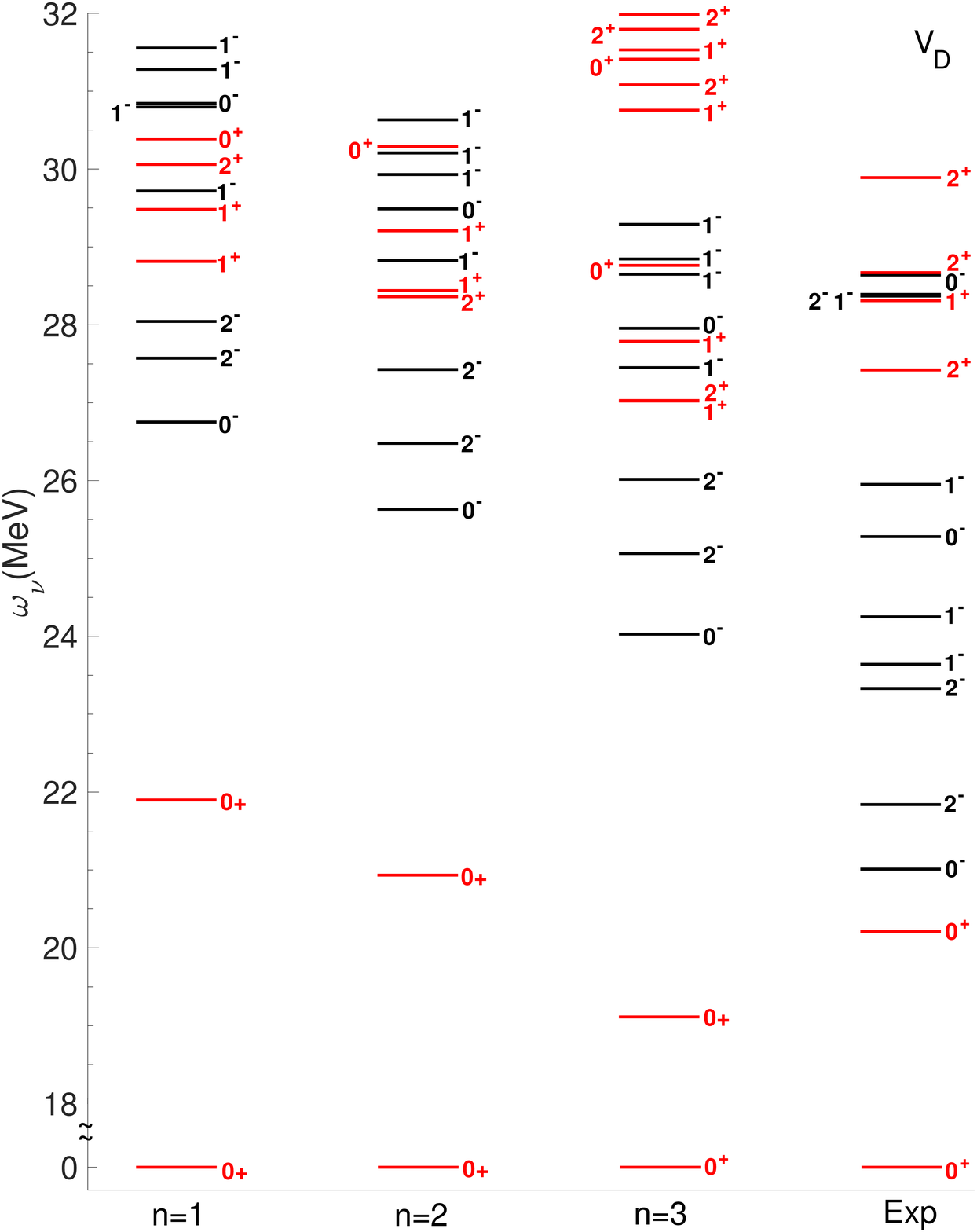}
\caption{(Color online) \label{fig4} The same as in Fig. \ref{fig3} for  V$_D$.}
\end{figure} 

We truncate the three-phonon subspace by including  all the states energies $E_{\alpha_2} +  E_\lambda < 100$ MeV to solve the final eigenvalue Eq. (\ref{eigfull}). Moreover, we keep only  the leading order term of the overlap matrix (\ref{D}) in computing the matrix elements ${\cal V}_{\alpha_1 \beta_3}$ (Eq. (\ref{couptwon})) and ${\cal V}_{\alpha_2 \beta_3}$ [Eq. (\ref{couponen})] which couple the three-phonon $ \ket{\beta_3} $ to the  one-phonon $\ket{\alpha_1 = \lambda} $ and  two-phonon $\ket{\alpha_2}$  states, respectively.

The diagonal approximation was tested in the restricted N$_{\rm max}=5$ HO space.  The deviations of the approximate energies from the exact ones range from  
$\sim 90$ keV to $\sim 250$ keV for V$_D$ and  from $\sim 30$ keV to $\sim 500$ keV for V$_S$. The only exception is represented by  the $2^-_1$ whose energy differs from the exact one by $\sim 1 $ MeV in the case of V$_S$. 

\begin{figure}[ht]
\includegraphics[width=\columnwidth]{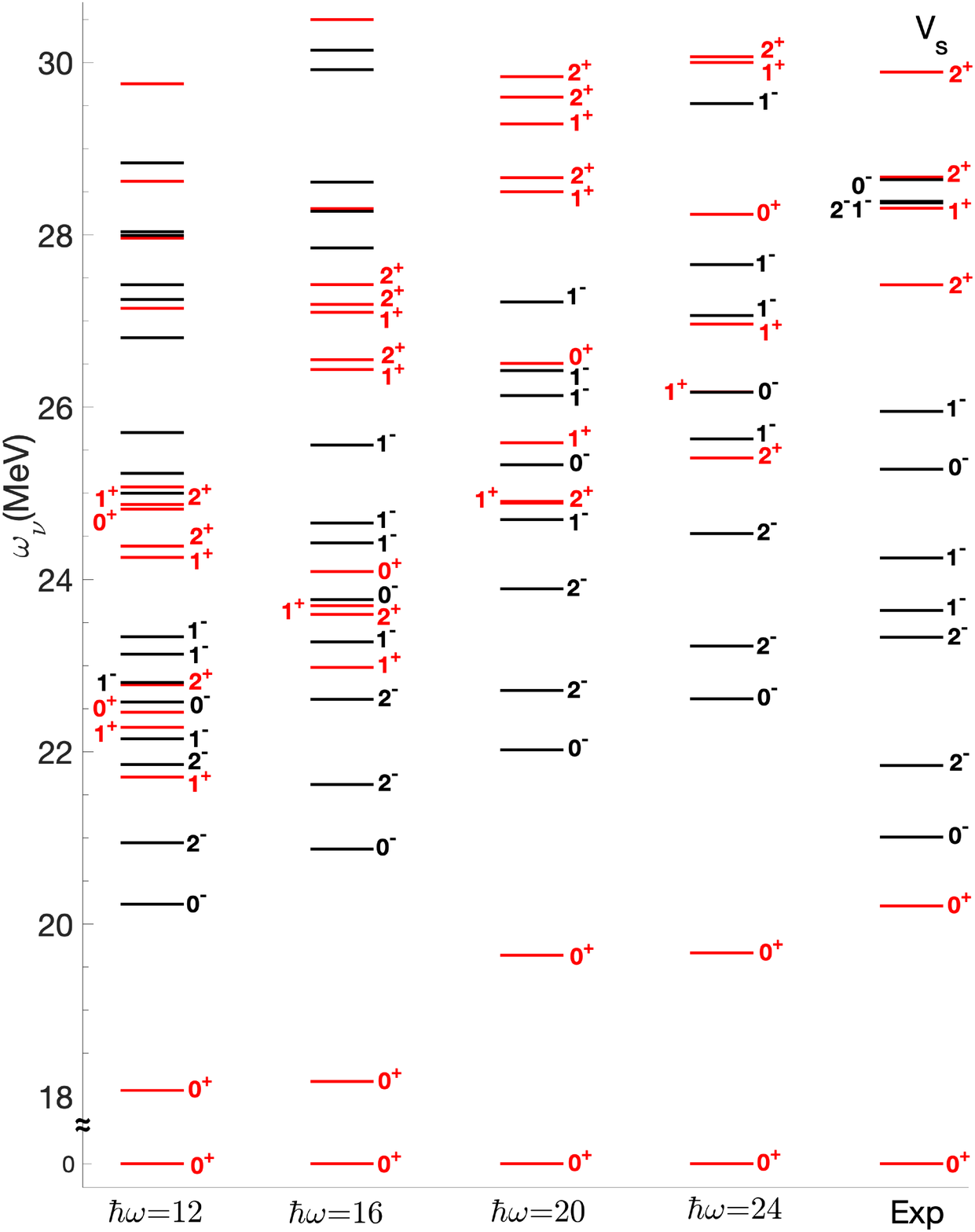}
\caption{(Color online) \label{fig5} Evolution of the V$_S$ spectrum with the HO frequency. The calculation is performed up to three phonons within a N$_{\rm max}=12$ HO space. The experimental data are from Ref. \cite{NNDC}.}
\end{figure} 

\begin{figure}[ht]
\includegraphics[width=\columnwidth]{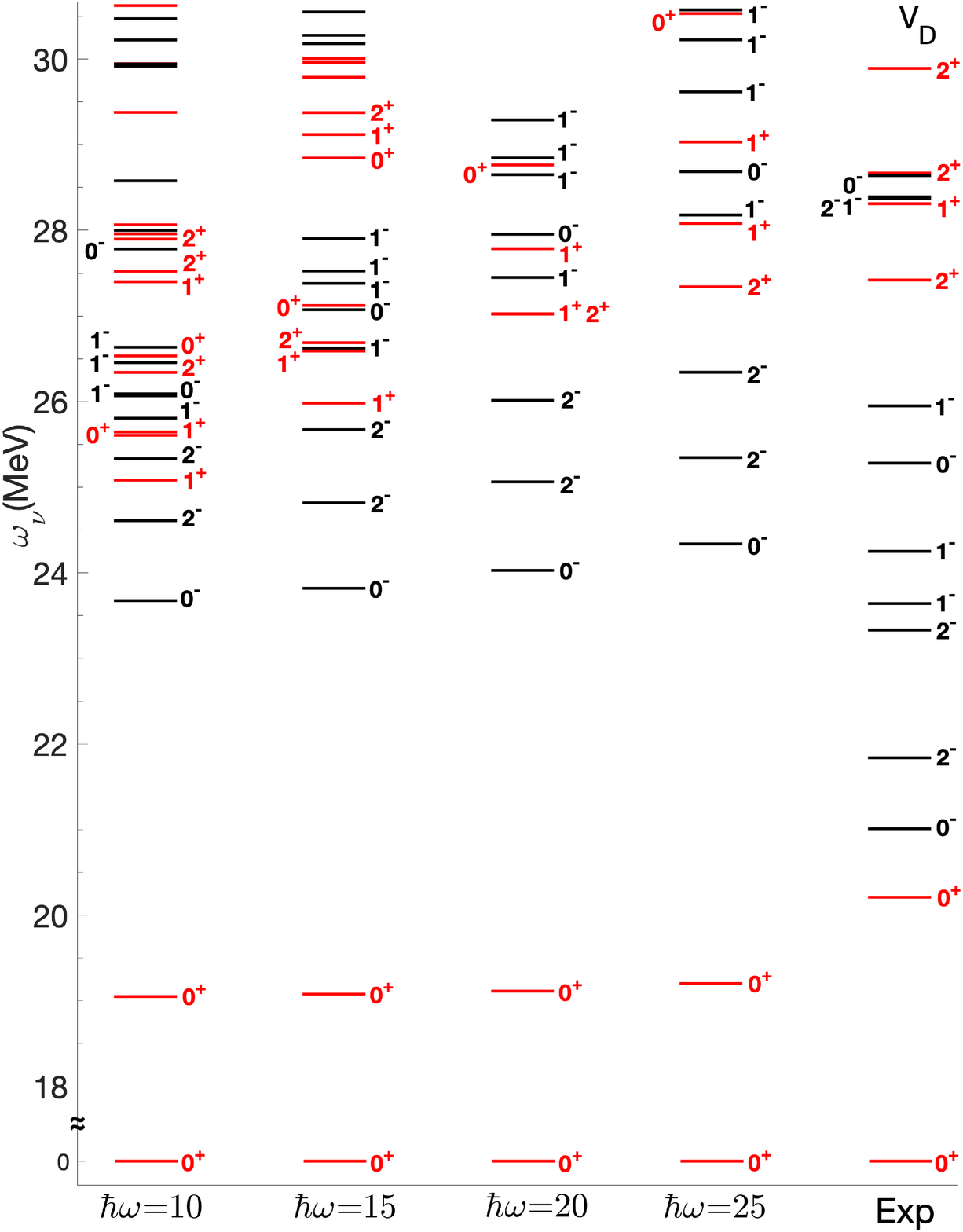}
\caption{(Color online) \label{fig6} The same as in Fig. \ref{fig5} for V$_D$.}
\end{figure}

\subsection{Ground state}
We have evaluated g.s. energies and proton radii for both potentials and studied their convergence properties with respect to HO frequencies and dimensions, up to $\hbar \omega =25$ MeV and N$_{max} =12$, respectively. We used intrinsic operators  in order to minimize the effect of the c.m. motion on the the HF states, as discussed in Ref. \cite{DEGREGORIO2021}.               
  
\subsubsection{Energy}
The g.s. energy produced by  V$_S$  is shared equally between HF and   two-phonon correlations (Fig. \ref{fig1}).  
The HF energy is almost insensitive to the HO frequency (Fig. \ref{fig1}) and reaches a stable value for N$_{\rm max} \geq 7$.  
The correlation energy, instead, depends appreciably on both frequency and dimensions and reaches convergence for N$_{\rm max} \geq 7$  and  $\hbar \omega \geq 20$ MeV. 
A small gap  ($\sim 1$ MeV) with the experiments remains. Most likely, it is due to the residual    three-body force, accounted for in creating the HF basis but  neglected in the multiphonon calculation.
Though determinant in   approaching the experimental binding energy, the two-phonon states account  only for $\sim$10\% of the wave function, which is dominated by HF (Table \ref{tab1a}).
  
In the case of V$_D$,   HF accounts almost entirely for energy (Fig. \ref{fig1}) and wave function (Table \ref{tab1b}).   
Nonetheless, the binding energy is reproduced just thanks to the  small contribution coming from the two-phonon correlations (Fig. \ref{fig1}). 
The convergence is reached for $\hbar \omega \geq 14$ MeV and N$_{\rm max} \geq 7$.

It is  worth mentioning that the g.s. energy is practically insensitive to the c.m. motion  whether we use V$_S$ or  V$_D$  in perfect agreement with the CC numerical proof that the use of an intrinsic Hamiltonian eliminates almost entirely the c.m. spurious admixtures from the g.s. \cite{Hagen09a,Hagen10a}.
For V$_S$ we get $E_{g.s.} =-27.042$ MeV  with c.m. admixtures and $E_{g.s.} = -26.866$ MeV without. For V$_D$ we obtain  $E_{g.s.}= - 28.274$ MeV with c.m. and $E_{g.s.} =-28.205$ MeV without. 

\subsubsection{Proton radius}
The proton square radius operator for $N=Z$ is  
\begin{eqnarray}
\label{rp2}
r_p^2 = \frac{1}{Z} \sum_{i=1}^Z \left(\vec{r}_i - \vec{R}_{cm}\right)^2 = \nonumber\\ 
= \frac{1}{Z} \left(1 - \frac{1}{A}\right)\sum_{i=1}^{Z} r_i^2 - \frac{1}{A^2}\sum_{i\neq j=1}^{A}\vec{r}_i \cdot \vec{r}_j.
\end{eqnarray} 
We have
\begin{equation}
\langle r_p^2 \rangle = \bra{\Psi_0} r_p^2 \ket{\Psi_0} = 
\langle r_p^2 \rangle_{HF} +  \langle r_p^2 \rangle_{corr}.
\end{equation}
The first is the HF term, while the second comes from the correlations and is given by 
\begin{eqnarray}
\langle r_p^2 \rangle_{corr} & = & \sum_{\alpha_n \alpha_{n'}'} {\cal C}^{0}_{\alpha_n}   {\cal C}^{0}_{\alpha_{n'}'}   \bra{\alpha_n} r_p^2 \ket{\alpha_{n'}'} = \nonumber\\
&=& \sum_{\alpha_2 \alpha_{2}'} {\cal C}^{0}_{\alpha_2}   {\cal C}^{0}_{\alpha_{2}'}   \bra{\alpha_2} r_p^2 \ket{\alpha_{2}'}, 
\end{eqnarray}
where  use of Eq. (\ref{Psifull}) has been made and
\begin{eqnarray}
\label{rptwo}
\bra{\alpha_2} r_p^2 \ket{\alpha_2'} =  \frac{1}{Z} \left(1 - \frac{1}{A}\right) \nonumber\\
\times \sum_{r s} \langle r \parallel r^2 \parallel s \rangle_p    
\langle \alpha_2 \parallel (a^\dagger_r \times  b_s)^0 \parallel  \alpha_2' \rangle.
\end{eqnarray}
It should be pointed out that only the two-phonon subspace contributes and that the two-body term of the square radius originating from the c.m. coordinates (\ref{rp2}) vanishes  because of the absence of c.m. spurious admixtures in the multiphonon wave function.  

The empirical value is extracted from the charge radius according to the formula \cite{Ekstr15}
\begin{equation}
\langle r_{ch}^2 \rangle =\langle r_p^2 \rangle + R_p^2 + \frac{N}{Z} R_n^2 + 
\frac{3 \hbar^2}{4 m_p^2 c^2},
\end{equation}
where $R_p = 0.8775(51)$fm, $R_n^2 = 0.1149(27)$fm$^2$, and $\frac{3 \hbar^2}{4 m_p^2 c^2} \sim 0.033$ fm$^2$.

As shown in Fig.  \ref{fig2},  HF yields almost the whole radius   for both potentials.
For N$_{\rm max} = 12$ , both  HF and total radii produced by V$_D$ are insensitive to any frequency. In the restricted  N=5 HO space, the convergence is reached for $\hbar \omega \geq 15$ MeV.
V$_S$ yields a HF radius roughly constant for all frequencies in both HO spaces.
The correlations do not alter the convergence properties for N$_{\rm max} = 12$. For N$_{\rm max} = 5$, instead, the convergence is reached  slowly for $\hbar \omega \geq 20$ MeV.
The radius obtained by V$_D$ almost coincides with the empirical value at the HF level but is shifted slightly upward by the correlations. 
V$_S$ produces a modest overestimation, incremented slightly by the correlations.

It is interesting to mention the effect induced by the c.m..
Let us consider the more sensitive  V$_S$ case.  If the c.m. is not removed,
the non diagonal matrix elements $\langle 0 \mid \vec{r}_1 \cdot \vec{r}_2 \mid \alpha_2 \rangle $  are non-vanishing. However, such a contribution is counterbalanced by a comparable enhancement of  the two-phonon matrix elements (\ref{rptwo}). 
Because of such a mutual cancellation, the radius remains practically unaffected, a further indication that the g.s. observables are insensitive to the c.m. motion.

\subsection{Spectra} 
\subsubsection{Impact of the c.m. motion}
In order to stress the vital importance of having a c.m. free spectrum, it is sufficient to analyze in more detail the results produced by the exact calculation using V$_S$ within the restricted N$_{max}$= 5 HO space  \cite{DEGREGORIO2021}.  Figs. \ref{figcm} and \ref{figwf} offer a vivid illustration of the dramatic impact of the c.m. motion on the excited states as we move from TDA ($n=1$) to the multiphonon space.  
The  TDA $1^-_1$  is practically entirely spurious and is nearly degenerate with the unperturbed HF g.s. state in perfect analogy with RPA. The close similarity between the two approaches, when both adopt a self-consistent single-particle or quasi-particle  basis, was discussed in Ref. \cite{Bianco14}. 

The spuriousness propagates among more and more states as the number of phonons increases and distorts dramatically the spectrum as well as the structure of the states. In fact, in addition to the dominantly spurious  one-phonon $1^-_1$,  which gets closer to the correlated g.s., an increasing number of states fall at too low energy. 
Most of them have two-phonon character and contain one or two spurious TDA  $1^-_1$ phonons (Fig. \ref{figwf}).  

It is worth noticing that all these states are not entirely spurious. Their spurious components are admixed with the physical ones of smaller amplitude. Moreover, spurious admixtures are present also in states with dominant c.m. free components. It is  therefore impossible to  disentangle the physical from the spurious states, hence the crucial role played by the SVD procedure.

\subsubsection{Convergence properties and comparative analysis}

The evolution of the V$_S$ spectrum with the $n$-phonon subspaces is similar for any HO frequency and dimensions. 

Let us consider N$_{\rm max} = 12$ and $\hbar \omega=20$ MeV (Fig. \ref{fig3}). In perfect analogy with the N$_{\rm max} = 5$ space, the  TDA spectrum falls in the experimental region but is too dense. Moreover, it should be pointed that the  levels are referred to the unperturbed HF  energy, which is $\sim 14$ MeV above the experimental value.  

The inclusion of the two-phonon subspace produces a large energy gap between excited and ground states and moves most levels above  the experimental region. In fact,  the coupling of the two phonons to HF [Eq. (\ref{couptwo})] is stronger than their coupling to  the one-phonon states  [Eq. (\ref{couponen})] and therefore  induces a strong depression of the g.s. only partly bridged by the downward shift of the excited states.

The coupling of the three phonons to one [Eq. (\ref{couponen})] and two [Eq. (\ref{couptwon})] phonons, treated in the diagonal approximation [Eq. (\ref{E3diag})], reduces drastically such a distance and brings the whole spectrum back to the experimental region.   

The  TDA  spectrum generated by V$_D$  is quite different (Fig. \ref{fig4}).  It is less dense but its levels are at too high energy. Also the evolution with the phonon number is different. No discontinuity is observed in going from one-phonon to two-phonon spaces.  Such a smooth behavior was largely expected given the small contribution ($\sim 2$ MeV) to the g.s. energy coming from the correlations. 
On the other hand, because of the minor impact of the multiphonon configurations, the levels get shifted downward smoothly but modestly and, therefore, remain at too high energies.
   
As shown in Figs. \ref{fig5} and \ref{fig6}, the trend of the level scheme with the HO frequency is similar for both potentials. It is too compressed and dense for low frequencies, due to the reduced distance between major shells.
For $\hbar \omega \geq 20$ MeV, the level density decreases in fair agreement with the experiments. The convergence with the frequency improves but not sufficiently. The  differences between the $\hbar \omega=20$ MeV and the $\hbar \omega=24$ MeV spectra are not negligible overall and seem to require additional HO shells for a satisfactory convergence, especially for V$_S$.

For a more detailed comparative analysis, it is appropriate to mention that the lowest seven negative parity  states plus the $0^+_1$ and $1^+_1$  undergo a nucleon decay while the  other levels, all  above $\sim 27$ MeV, undergo a deuteron($D$) decay \cite{NNDC}. The first should be put in correspondence with states having a dominant one-phonon component, while the second levels should be associated to states having a dominant two-phonon structure.

From  Figs. \ref{fig3} and \ref{fig4}, we observe for both V$_S$ and V$_D$ spectra a one-to-one correspondence between the first seven theoretical and experimental  negative-parity levels.    
They are, respectively, $\sim 1$ and $\sim 3$ MeV above.   These states have a dominant one-phonon character  (Tables \ref{tab1a} and \ref{tab1b}) and a p-h  content roughly compatible with the decay of the corresponding  experimental levels  (Table \ref{tab2a}).

\begin{table}  
\caption{\label{tab1a} Energies and $n$-phonon composition of the lowest states 
computed using V$_{S}$ for $\hbar\omega=20$ MeV and N$_{\rm max}=12$. $W^{\nu}_n=\sum_{\alpha_n} \mid {\cal C}^{\nu}_{\alpha_n}\mid^2$ [see Eq. (\ref{Psifull})].}
\begin{ruledtabular}
\begin{tabular}{cccccc}
$J^{\nu}$&$E^{\nu}$&$W^{\nu}_0$&$W^{\nu}_1$&$W^{\nu}_2$&$W^{\nu}_3$\\
\hline
$0^+_1$&    0.000 &  0.893  &   0.007   & 0.099   &  0.000\\
$0^+_2$&   19.639  &    0.005   &  0.807  &   0.114   &  0.074\\
$0^-_1$&   22.023   &  0.000  &  0.833   & 0.099  &  0.068\\
$2^-_1$&  22.711  &   0.000  &  0.843   & 0.090  &  0.068\\
$2^-_2$&   23.892  &   0.000 &   0.863 &  0.066  &  0.071\\
$1^-_1$ & 24.694   &  0.000   & 0.859   & 0.069   & 0.071\\
$2^+_1$ & 24.885   &  0.000   & 0.832  &  0.102   & 0.066\\
$1^+_1$   &  24.907  &   0.000  &  0.860 &   0.068 &   0.072\\
$0^-_2$  &  25.331  &   0.000   & 0.825 &  0.105   & 0.070\\
$1^+_2$ &   25.587   &  0.000  &  0.867  &  0.059  &  0.073\\
$1^-_2$   & 26.135   &  0.000   & 0.812  &  0.122  &  0.066\\
$1^-_3$&  26.426    & 0.000 &   0.869  &  0.058  &  0.073\\
$0^+_3$  & 26.508  &   0.000  &  0.878  &  0.046 &   0.075\\
$1^-_4$   &  27.220  &   0.000 &   0.857  &  0.072&    0.071\\
$1^+_3$  &  28.500  &   0.000 &   0.851   & 0.078   & 0.071\\
$2^+_2$   &  28.664  &   0.000  &  0.869 &   0.059 &  0.072\\
$1^+_4$   &   29.286   &  0.000  &  0.855  &  0.073&   0.071\\
$2^+_3$  &  29.600  &   0.000 &   0.873  &  0.054   & 0.073\\
$2^+_4$   &   29.835  &   0.000   & 0.840 &   0.091  &  0.068\\
$2^+_6$ &     37.780 &    0.000   & 0.004&    0.970   & 0.026\\
 \end{tabular}
\end{ruledtabular} 
\end{table} 

\begin{table}  
\caption{\label{tab1b} The same as Table \ref{tab1a} for V$_D$.}
\begin{ruledtabular}
\begin{tabular}{cccccc}
$J^{\nu}$&$E^{\nu}$&$W^{\nu}_0$&$W^{\nu}_1$&$W^{\nu}_2$&$W^{\nu}_3$\\
\hline
$0^+_1$&  0.000  &  0.975 &  0.000 &   0.024  &  0.000\\
$0^+_2$&19.112  &   0.001 &   0.863 &   0.111    &0.026\\
$0^-_1$& 24.028   &  0.000   & 0.862 &   0.116  &  0.022  \\
$2^-_1$&  25.063 &    0.000   & 0.876  &  0.106  &  0.018\\
$2^-_2$&26.015   &  0.000   & 0.890   & 0.092   & 0.018\\
$1^+_1$&27.021  &   0.000  &  0.890   & 0.092   & 0.018 \\
$2^+_1$  &27.027  &   0.000&   0.847  &  0.136 &   0.017 \\
$1^-_1$ &27.450    & 0.000   & 0.878  &  0.104  &  0.018 \\
$1^+_2$& 27.787 &   0.000  &  0.896 &   0.086  &  0.018 \\
$0^-_2$& 27.954  &   0.000   & 0.824  &  0.152   & 0.023  \\
$1^-_2$& 28.649   &  0.000   & 0.840 &   0.143   & 0.017 \\
$0^+_3$& 28.763  &   0.000 &   0.888  &  0.091  &  0.022 \\
$1^-_3$& 28.845   &  0.000  &  0.863  &  0.119   & 0.018 \\
$1^-_4$& 29.289   &  0.000    &0.877  &  0.105  &  0.018\\
$1^+_3$   &  30.756  &   0.000  &  0.865    &0.117   & 0.018  \\
$2^+_2$   &31.082    & 0.000 &   0.884  &  0.099 &   0.017 \\
$0^+_4$   &31.411  &  0.000  &  0.797 &  0.176 &   0.027  \\
$1^+_4$  &31.531   &  0.000  &  0.872   & 0.111   & 0.017 \\
$2^+_3$  &31.793   &  0.000  &  0.856   & 0.127    &0.017 \\
$2^+_4$   & 31.981 &    0.000 &   0.889   & 0.093   & 0.017 \\
 $2^+_6$ & 38.102   &  0.000 &   0.006   & 0.968  &  0.025\\
 \end{tabular}
\end{ruledtabular} 
\end{table} 

\begin{table}  
 \caption{\label{tab2a} Proton ($\tau=p$) and neutron ($\tau=n$) p-h   weights $W_\tau = \sum_{ph} |c^{\tau}_{ph}|^2$ of the one-phonon (TDA) components of the lowest excited V$_S$ and V$_D$ states  compared with the proton (\%$p$) and neutron (\%$n$) decay modes of the experimental levels \cite{TILLEY1992}. }
 \begin{ruledtabular}
 \begin{tabular}{ccccccc}
$J^{\pi}_\nu$&$ W^{V_{S}}_p$&$W^{V_{S}}_n$&$ W^{V_{D}}_p$&$W^{V_{D}}_n$&\%$p$ &\%$n$ \\
\hline
$0^+_ 2 $ &54 &46& 47&53&100&0\\
$ 0^-_1  $&  59    & 41& 55     &45 &76&24   \\
$2^-_1   $&  91   &9&  87  &13&63& 37\\
$2^-_2  $&   10   &90& 12     &88&53&47\\
$ 1^-_1  $&59    & 41 &  62  & 38 &55&45 \\
$ 0^-_2 $& 41    & 59& 46    & 54&52&48    \\
$ 1^-_2  $&  64   & 36 &83     &17  &50&47  \\ 
$ 1^-_3  $&   41 & 59& 22   &78 &52&48    \\  
$ 1^+_1 $&96     &4&82     &18&48&47 \\
\end{tabular}
\end{ruledtabular} 
\end{table} 

Also the lowest $0^+$  of both computed spectra has a one-phonon structure and can be related to the experimental $0^+$.   
However, its p-h content does not match the decay products of the experimental level.
It is problematic to identify, among the several theoretical $1^+$ levels occurring in both V$_S$ and V$_D$ spectra, the counterpart of the experimental $1^+$. 

The theoretical spectra, especially the  one generated by V$_S$, contain additional low-lying positive parity levels plus a $1^-$. They have a one-phonon  character  (Tables \ref{tab1a} and \ref{tab1b})  and cannot  be associated to  the three experimental  $2^+$ and the $0^-_3,1^-_4,2^-_3$ triplet,  all undergoing a $D$ decay.
The occurrence of these theoretical low-lying positive-parity one-phonon intruders may be traced back to the rather small  $(1s,0d)-(0s)^{-1}$ p-h HF energies which,  in turn, generate low TDA levels. We do not have any obvious explanation for the $1^-$ one-phonon intruder.

The two-phonon states, of both positive and negative parity, which can be associated to the $D$-decaying levels are  at too high energies ($\sim 38$ MeV or above).  
There are, however, mechanisms for pushing them down.  One may consist in enlarging the HO space.
Another one is suggested by Eq. (\ref{couptwon}) which shows that  the $(n+2)$-phonon to $n$-phonon coupling is proportional to the strong coupling of two phonons to the HF vacuum.  We have seen already that such a coupling is responsible for the strong impact of the  three phonons on the one-phonon levels.
It is therefore natural to expect an analogous effect of the four phonons on the two-phonon states.
\begin{figure}[ht]
\includegraphics[width=\columnwidth]{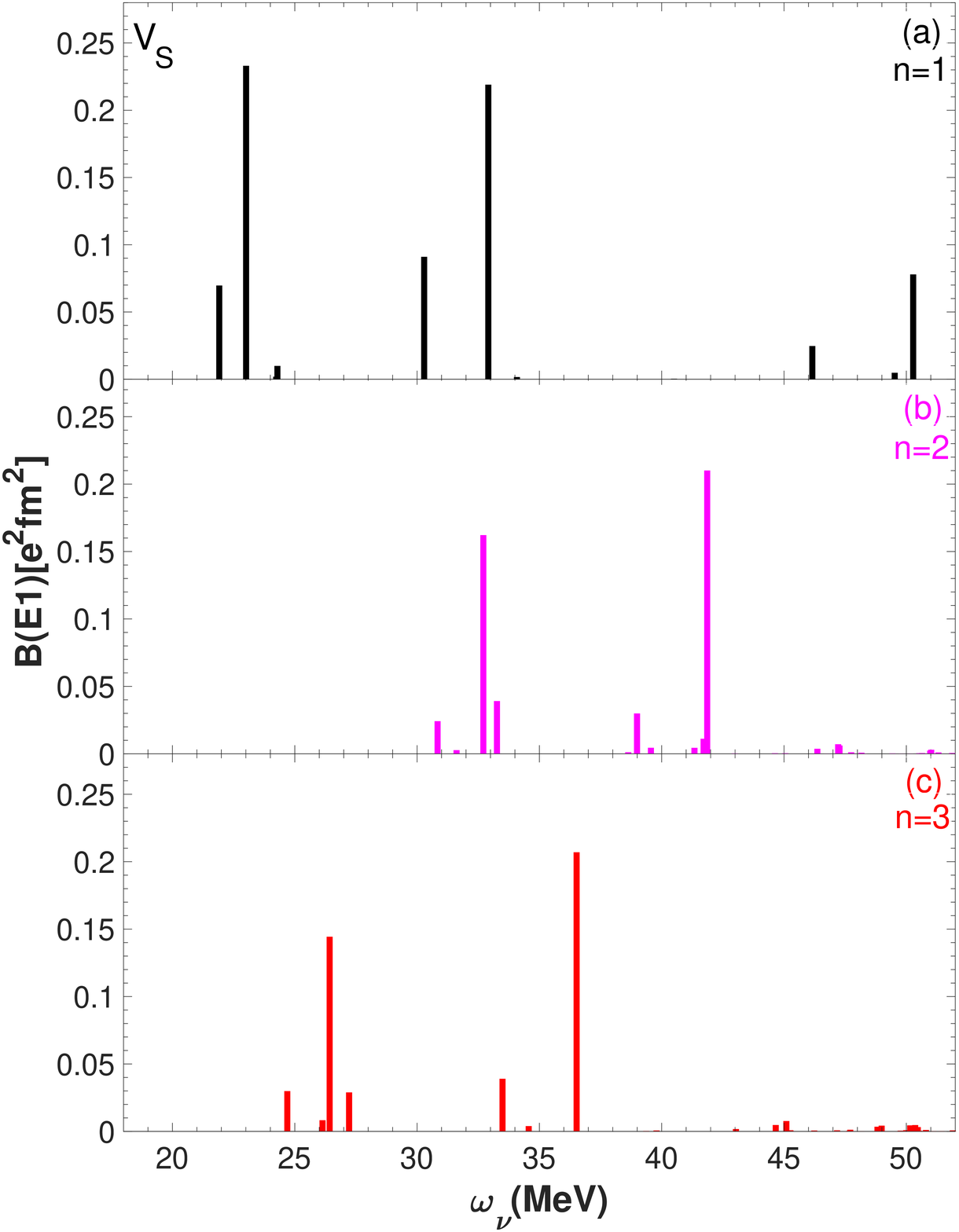}
\caption{(Color online) \label{fig7}  $E1$ spectra computed using V$_S$ and 13 major shells within spaces including up to $n=1,2,3$ phonons.}
\end{figure}
  
\begin{figure}[ht]
\includegraphics[width=\columnwidth]{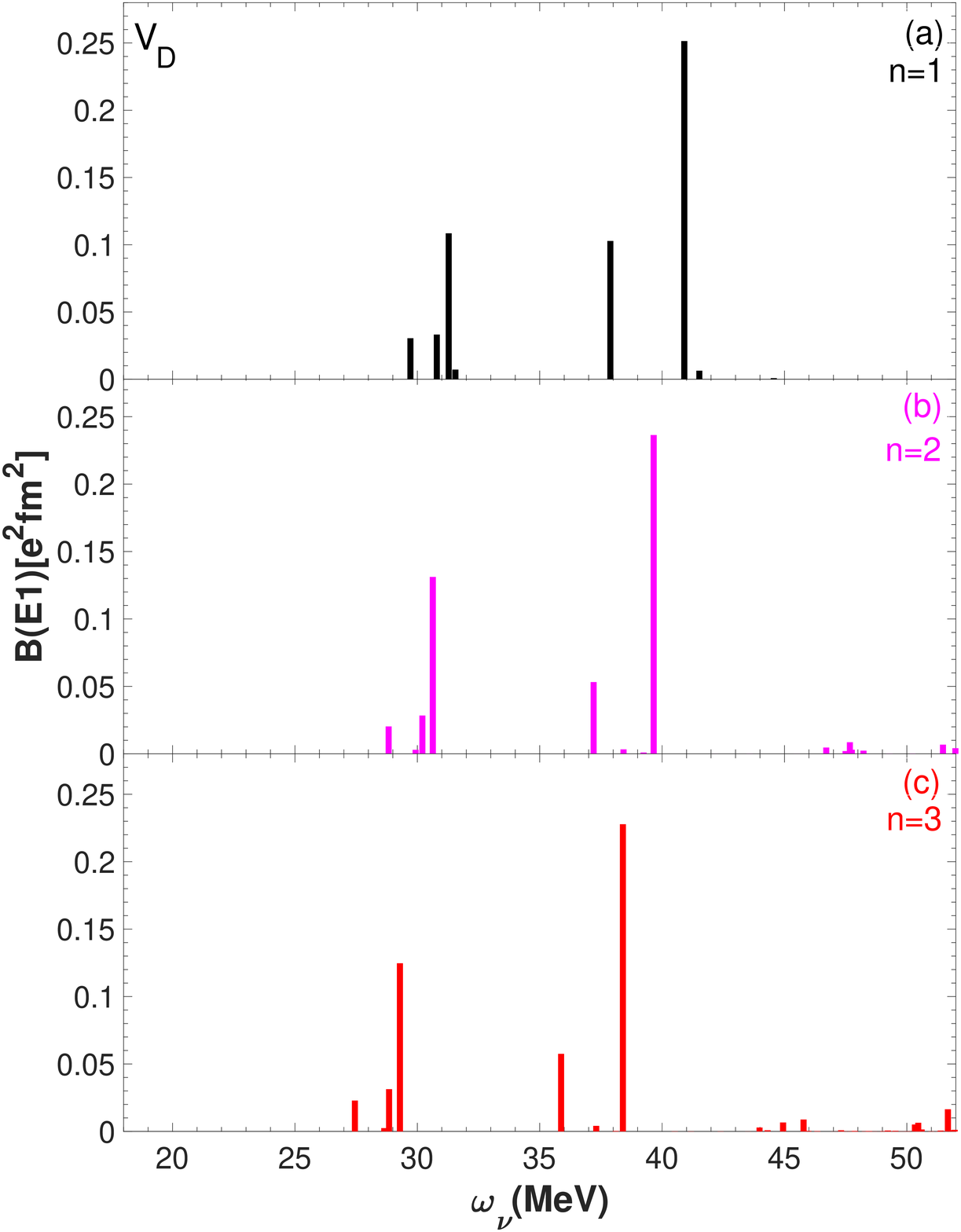}
\caption{(Color online) \label{fig8} The same as in Fig. \ref{fig7} for  V$_D$.}
\end{figure}

In order to obtain a rough estimate of the impact of such a coupling,  we make the simplifying assumption that the four-phonon states are composed of two non interacting two-phonon states so that   
\begin{equation}
E_{\beta_4} \simeq E_{\beta_2} +  E_{\beta_2'}.
\end{equation}
We truncate the subspace by imposing the constraint $E_{\beta_2} +  E_{\beta_2'} < 150$ MeV. Furthermore, we compute the coupling [Eq. (\ref{couptwon})]  by keeping only the leading order term of the overlap matrix ($\braket{ (\alpha_2 \times \alpha_2')^\beta}{\beta _4} \sim \braket{(\alpha_2 \times \alpha_2')^\beta}{(\beta_2 \times \beta_2')^\beta} \sim \delta_{\alpha_2 \beta_2}  \delta_{\alpha_2' \beta_2'}$).

 The two-phonon levels move from $\sim 38$ to $\sim 33$ MeV, appreciable but not sufficient to fill the gap with the experiments. We could enlarge the four-phonon subspace for a further shift. For our purpose, however,
 it is enough to show that  the coupling to four phonons is necessary for an exhaustive description of the full spectrum.
    
\begin{figure}[ht]
\includegraphics[width=\columnwidth]{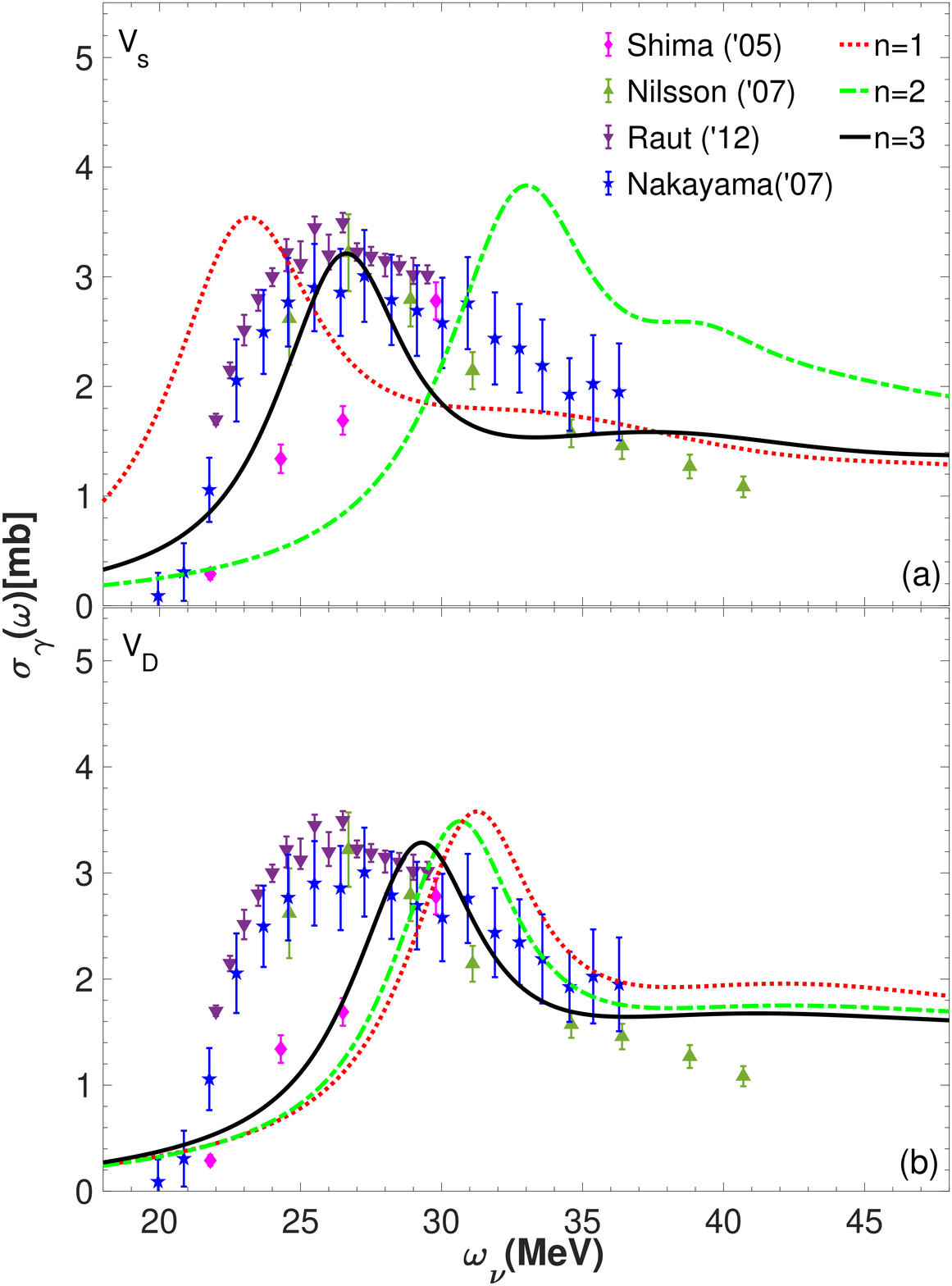}
\caption{(Color online) \label{fig9} Theoretical GDR cross section computed using   V$_S$ (a) and V$_D$ (b)  in spaces including up to $n=1$ (dashed line) $n=2$ (dotted line), and $n=3$ (continuous line). The calculation is performed up to three phonons within a N$_{\rm max}=12$ HO space for $\hbar\omega=20$ MeV.  The experimental data are taken from Refs. \cite{Shima05,Nilsson07,Raut12,Nakayama07}, assuming, following Ref. \cite{Schuster2015},  $\sigma_\gamma (\omega)\approx  2\sigma_{\gamma,n}(\omega)$  for the data of Ref. \cite{Raut12} and $\sigma_\gamma (\omega)\approx  \sigma_{\gamma,p}(\omega)+\sigma_{\gamma,p}(\omega+0.5$ MeV)  for those of Ref. \cite{Nilsson07}.} 
\end{figure}

\begin{figure}[ht]
\includegraphics[width=\columnwidth]{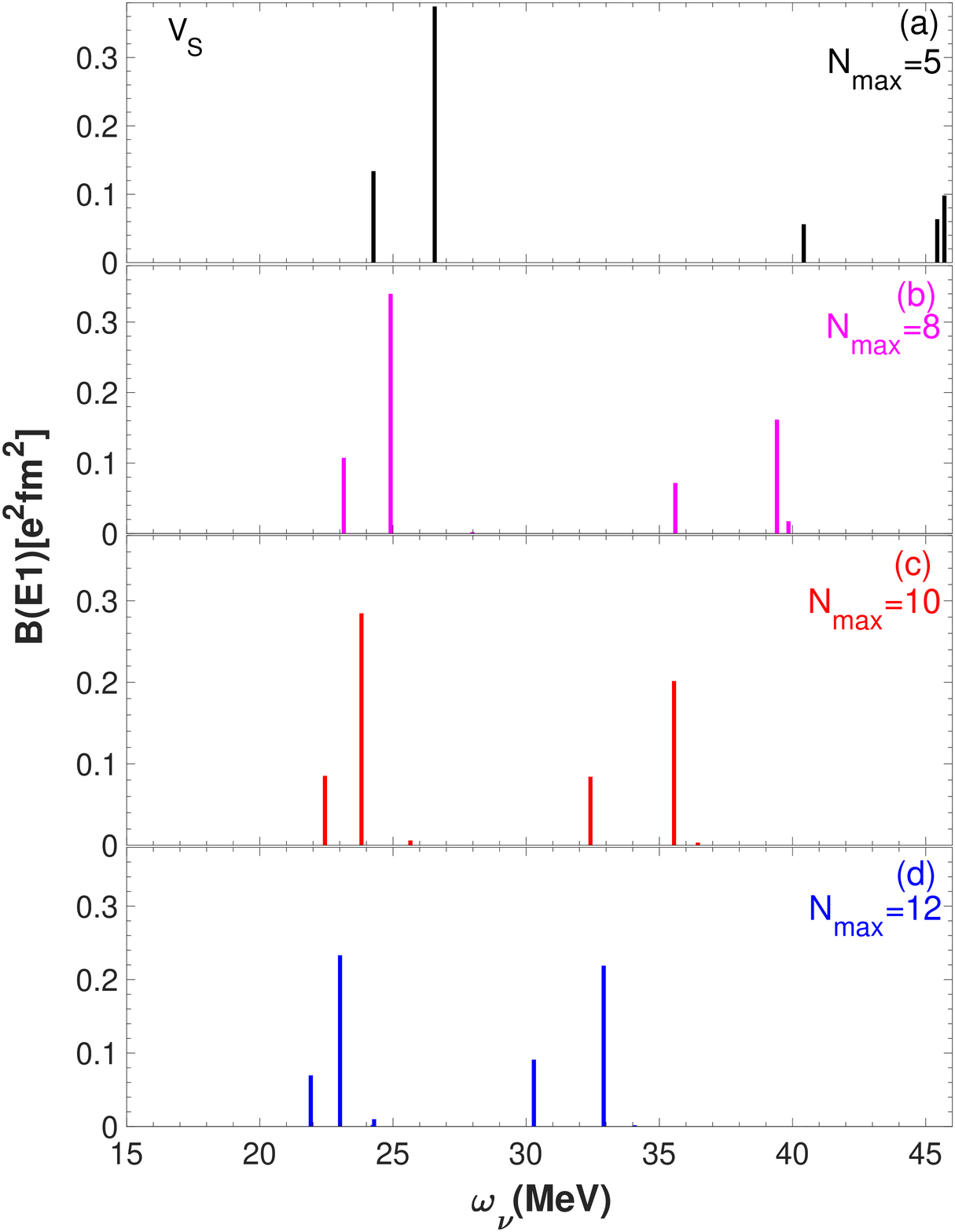}
\caption{(Color online) \label{fig10}  TDA $E1$ spectra computed using V$_S$  versus the HO dimensions N$_{max}$ for $\hbar\omega=20$ MeV.}
\end{figure} 
 
\subsection{Dipole transition amplitudes and giant resonance}
The absorption electric dipole ($E1$) cross section is given by
\begin{eqnarray}
\sigma (E1) &=&  \int_{0}^{\infty} \sigma (E1, \omega)   d\omega 
\nonumber\\
&=&\frac{16 \pi^3}{9\hbar c} \int_{0}^{\infty} \omega {\cal S}(E1, \omega) d\omega, 
 \label{sigma}
\end{eqnarray}
where ${\cal S}(E1, \omega)$ is the   strength function   
\begin{equation}
{\cal S}(E1, \omega) = \sum_{\nu } B_{\nu}  ( E1)\,\delta (\omega -\omega_{\nu }).
\label{Strength}
\end{equation}
The reduced strength 
 \begin{equation}
B_\nu (E1)=  \mid \langle  \Psi_{\nu}  \parallel  {\cal M}(E1) \parallel   \Psi_{0} \rangle \mid^2
\end{equation} 
is determined by the  g.s. transition to the $\nu_{th}$ final state of energy  $\omega_\nu = {\cal E}_\nu - {\cal E}_{0}$    
\begin{eqnarray}
\label{ME1}
 \langle \Psi_{\nu} \parallel  {\cal M}(E1) \parallel   \Psi_0  \rangle = \nonumber\\
 \sum_{\alpha_n \alpha_{n'}} 
 {\cal C}^{0}_{\alpha_n} {\cal C}^{\nu}_{\alpha_{n'}}
  \langle \alpha_{n'}\parallel {\cal M} (E1) \parallel\alpha_n \rangle,
  \end{eqnarray}
having made use of  Eq. (\ref{Psifull}) for the wavefunctions. 

The  electric dipole ($\lambda = 1$) operator  has the standard form ($\tau=p,n$)
\begin{equation}
\label{Mlaphop}
{\cal M}(E \lambda \mu) = \frac{1}{[\lambda]^{1/2}}
\sum_{(rs)_\tau} e_\tau \langle r \parallel r^\lambda Y_\lambda  \parallel   s \rangle \Bigl[a_r^\dagger \times b_s \Bigr]^\lambda_ \mu 
\end{equation}
with proton and neutron bare charges $e_p=e$ and $e_n=0$.

In the present calculations  we replace, as common practice, the $\delta$ function appearing in the strength function [Eq. (\ref{Strength})] with a Lorentzian of width $\Delta$.

Our procedure differs from the one adopted in {\sl ab-initio} calculations  \cite{Quaglioni2007,Bacca2014,Schuster2015} which exploit the Lorentz transform \cite{Efros94} and 
do not require the explicit determination of all $1^-$ states as in our case. Thus, position and shape of our cross section depend critically on the positions of the $1^-$ levels and on the distribution of $E1$ strength among them.  
 
The $E1$ g.s. transition strength is ultimately due to the  two transitions from the $0s_{1/2}$  to the $0p_{3/2}$  and $0p_{1/2}$ single proton HO states. 
In HF, the strength spreads over all p-h states containing the $0p_{3/2} - 0s_{1/2}^{-1}$ and $0p_{1/2}-0s_{1/2}^{-1}$ HO p-h states.

In TDA, the V$_S$ strength gets distributed almost equally among two groups falling in the regions  $20$-$25$ and $30$-$35$ MeV (Fig. \ref{fig7}).  Both are composed of a short peak and a high peak.
The short ones are promoted by the $0s_{1/2} \rightarrow 0p_{3/2}$ transition. The high peaks are due to  the $0s_{1/2} \rightarrow 0p_{1/2}$ transition. 
A residual strength is located in the $45$-$50$ MeV region. 
An analogous $E1$ spectrum was obtained within a RPA approach using the same potential V$_S$ \cite{Wu2018}.
The multiphonon states, while fragmenting completely the strength located at $45$-$50$ MeV, cause an overall damping of the other peaks  with the exception of the one at  $\sim 36.5$ MeV. However, the two-branch structure persists
and the distance between the two groups is unchanged.

The  V$_D$ strength splits also into two branches. They are only shifted upward in energy. Such a structure remains unchanged once the two and three phonons are included (Fig. \ref{fig8}).

The two-branch profile persists  for both potentials even if we change the frequency. 
Therefore, it came out to be impossible to try to reproduce the shape of the experimental cross section by using a single width for all levels. The best we have been able to do is to choose two widths.  By doing so, we obtain for both potentials a single hump which only roughly approaches the shape of the experimental cross section (Fig. \ref{fig9}).  

This result is different from the one obtained in the small space  N$_{\rm max}=5$, where it was possible to obtain a single hump by using a single width \cite{DEGREGORIO2021}. 
This difference can be understood if we observe the evolution of the $E1$ strength as the HO dimensions increase. 
As shown in Fig. \ref{fig10}, for N$_{\rm max} =5$,  the V$_S$ strength is concentrated almost entirely into two close TDA peaks. The one at $\sim 24.3$ MeV collects a strength $B(E1) \sim  0.13$ $e^2fm^2$ 
and is due mainly to the $0s_{1/2} \rightarrow 0p_{3/2}$ transition. The other at $\sim 26.6$ MeV is much stronger ($B(E1) \sim  0.37$ $e^2fm^2$)  and arise from the $0s_{1/2} \rightarrow 0p_{1/2}$ transition. 
A marginal strength appears at high energy, $\sim 45$ MeV, and does not interfere with the low-energy hump.
The multiphonon states causes only a damping and an upward shift. It was therefore possible to approach the experimental cross section by using a single Lorentzian width $\Delta= 10$ MeV.

However, as N$_{max}$ increases, the high energy peaks move from $\sim 45$ MeV downward and tend to approach the low-lying branch, whose energy remains almost constant. Moreover, it becomes stronger  at the expense of the low-energy transitions.  In fact, for N$_{\rm max}=12$, the TDA strength gets distributed among two doublets of comparable strength.  With respect to the small space, the energy separation between them is much smaller  but still large ($\sim 10 $ MeV). The coupling to three phonons depletes further the low energy peaks in favor of the second ones, but leaves unchanged the separation between the two regions (Fig. \ref{fig7}). Hence the impossibility of getting a one-hump profile by a single width.

The above analysis, however, suggests how to reach our goal.  If we enlarge further the HO space,  following the trends illustrated in Figs. \ref{fig7} and \ref{fig10}, the second  branch is expected to go down in energy and to collect most of the strength at the expense of the first one, especially once the three-phonon states come into play. Therefore, it should be possible to approach position and shape of the experimental cross section by using a single Lorentzian width. Also the full spectrum should benefit from moving to a larger space.
 
\section{Conclusions}
The different characteristic of the two potentials V$_D$ and V$_S$ have visible effects on the bulk properties of $^4$He. V$_D$ privileges HF, which accounts almost entirely for the g.s. energy and proton radius and therefore promotes  a fast convergence versus HO frequency and dimensions. Such a fast convergence is due to   the softening of the potential induced by SRG. In the case of V$_S$, instead, HF and two-phonon correlations contribute on equal footing to the energy.  Consequently, convergence is reached in a more restricted range of frequencies and sufficiently large dimensions. Such a poor convergence was expected since a bare $NN$+$NNN$ force was used.

The distinct peculiarities of the two potentials get manifested also in the spectrum. 
Because of the dominant role played by HF, the level scheme produced by V$_D$ evolves smoothly as we move from the one-phonon to the three-phonon space.
The multiphonon states have little impact and do not alter significantly the TDA spectrum. Thus, the resulting levels remain too high with respect to the  experimental  ones
for any HO frequency. In fact, though convergence is not reached, the variations with $\hbar \omega$ are small and tend to enlarge the gap as we increase the frequency. We do not expect a significant improvement even if we enlarge the HO space, given the fast convergence  of HF versus N$_{max}$ and the marginal role played by the multiphonon configurations.  

In the case of V$_S$,  the multiphonon states play an essential role. The spectrum undergoes dramatic changes in going from $n=1$ to $n=2$ and, then, from $n=2$ to $n=3$ phonon spaces. Only once the three-phonon states are included, is it possible to establish a satisfactory correspondence between the dominantly one-phonon states and the nucleon decaying levels. No convergence versus the HO frequency is observed. The best proximity to the experiments is obtained for frequencies around $\hbar \omega= 20$ MeV. Given the important role played by the correlations, we should expect further improvements from enlarging the HO space. The dominantly two-phonon states remain at too high energy with respect to the $D$-decaying levels. However, we have shown that  the gap can be drastically reduced by their coupling to four phonons.

The $E1$  V$_S$ and V$_D$ spectra present some analogies but also important differences.   In both cases, the strength is distributed mostly among two groups of levels. However, the V$_D$ spectrum is at too high energy and unaffected by the multiphonon configurations. These, instead,  are quite effective in the V$_S$ spectrum. In fact, the strength tends to shift from the low to the high energy group of levels as we move from the $n=1$ to the $n=3$ space. Another important feature is its sensitivity to the HO space dimensions. The high-energy peaks become more prominent and tend to move down toward the low-lying group of levels as we enlarge the HO space. 

This strength redistribution has important  impact on the $E1$ giant resonance.
Since the two groups of levels are still too far apart for N$_{max}$ = 12, we had to use a small and a high Lorentzian width in order to reproduce roughly position and shape of the experimental cross section. We expect that, by enlarging further the HO space, the second dominant group of levels should approach closely the low-lying peaks and, eventually, merge with them. In such a case, a more faithful description of the giant resonance  profile should be achieved by enveloping the peaks with a single width. Should this recipe fail, we must conclude that  the states constructed out of a bound single-particle basis are not adequate for describing the experimental $1^-$ excitations lying in the deep continuum. The only remaining alternative would be to resort to the Lorentz transform method.

In summary, the V$_D$ potential has the nice property of privileging HF, thereby promoting a fast convergence versus frequency  and dimensions of the HO space. However, it leaves little room for improving the agreement with the experiments through the correlations. These, instead, are essential when V$_S$ is used and, though only for specific frequencies,  provide a better description of the spectroscopic properties at the cost of increasing more the dimensions of the HO space.  We are confident, however, that  even a modest expansion of the space adopted here will promote a significant progress in  the description of spectrum and giant resonance.

\begin{acknowledgments}
We thank Petr Navr\'{a}til, Mark A. Caprio, Patrick J. Fasano, and Jakub Herko for having provided the matrix elements of the NNLO$_{\rm{sat}}$  and Daejeon16 potentials. This work was partly supported by the Czech Science Foundation (Czech Republic), P203-19-14048S and by the Charles University Research Center UNCE/SCI/013. F.K. and P.V. thank the INFN for financial support. 
Computational resources were provided by the CESNET LM2015042 and the CERIT Scientific Cloud LM2015085, under the program "Projects of Large Research, Development, and Innovations Infrastructures". G. De Gregorio acknowledges the support by the funding program "VALERE" of the Universit$\grave{a}$ degli Studi della Campania "Luigi Vanvitelli".
\end{acknowledgments}
\bibliographystyle{apsrev}
\bibliography{Hefour}
\end{document}